\documentclass[conference,letterpaper]{IEEEtran}

\usepackage[utf8]{inputenc} 
\usepackage[T1]{fontenc}
\usepackage{url}
\usepackage{ifthen}
\usepackage{cite}
\usepackage[cmex10]{amsmath} %

\usepackage{amsthm}
\usepackage[inline]{enumitem}
\newtheorem{lemma}{Lemma}
\newtheorem{remark}{Remark}

\newtheorem{auxiliary code}{Auxiliary Code}

\newtheorem{definition}{Definition}

\newtheorem{strategy}{Dispersal Protocol}
\usepackage{algorithm}
\usepackage[noend]{algpseudocode}
\algrenewcommand\algorithmicindent{0.8em}%

\def \tG {\mathcal{G}}
\def \tC {\mathcal{C}}
\def \tS {\mathcal{S}}
\def \tnei {\mathrm{neigh}}
\def \tTG {\widehat{\mathcal{G}}}
\def \tL {\mathcal{L}}
\def \tK {\mathcal{K}}
\def \tcn {c}
\def \tCg {\mathcal{V}^{gr}}
\def \tr {R}

\usepackage{tikz}
\usepackage{graphicx}
\usetikzlibrary{positioning}
\usepackage[font={small}]{caption}
\usepackage{subcaption}
\usepackage{multirow}
\usepackage[T1]{fontenc}

\newcommand\deb[1]{{\color{black}#1}}
\newcommand\debb[1]{{\color{black}#1}}
\newcommand\debbb[1]{{\color{black}#1}}
\newcommand\debbbb[1]{{\color{black}#1}}

\def \tgreedy {{\fontfamily{qcr}\selectfont greedy-size}}
\def \tPEG {{\fontfamily{qcr}\selectfont PEG}}

\begin{document}
\title{ \deb{Communication-Efficient LDPC Code Design for Data Availability Oracle in Side Blockchains\vspace{-0.5cm}}}

\author{\IEEEauthorblockN{Debarnab Mitra, Lev Tauz and Lara Dolecek}
\IEEEauthorblockA{Department of Electrical and Computer Engineering, University of California, Los Angeles, USA\\
email: debarnabucla@ucla.edu, levtauz@ucla.edu, dolecek@ee.ucla.edu}
\vspace{-1cm}}

\maketitle

\begin{abstract}
A popular method of improving the throughput of blockchain systems is by running smaller side blockchains that push the hashes of their blocks onto a trusted blockchain. Side blockchains are vulnerable to \emph{stalling attacks} where a side blockchain node pushes the hash of a block to the trusted blockchain but \debbb{makes the block unavailable to other side blockchain nodes}. Recently, Sheng \emph{et al.} proposed a data availability oracle based on LDPC codes and a data dispersal protocol as a solution \debb{to the above problem}. \debbb{While showing improvements, the codes and dispersal protocol were designed disjointly which} may not be optimal in terms of the communication cost associated with the oracle.
 In this paper, we provide a tailored dispersal protocol and specialized LDPC code construction based on the Progressive Edge Growth (PEG) algorithm, called the \emph{dispersal-efficient} PEG (DE-PEG) algorithm, aimed to reduce the communication cost associated with the new dispersal protocol. Our new code construction reduces the communication cost \debbb{and, additionally,} is less restrictive in terms of system design.
\end{abstract}
\vspace{-0.05cm}
\vspace{-0.37cm}
\section{Introduction}
\vspace{-0.2cm}

Side blockchains, e.g., \cite{CharitySideBlockchain,Binance,Blockcert,BuildingBlocks}, are a popular method of improving the transaction throughput of blockchain systems where a single trusted blockchain supports a large number of \debbb{side (smaller) blockchains} by storing the block hashes of the side blockchains in their ledger \cite{AceD}. 
Systems that run side blockchains are vulnerable to a form of data availability attack \cite{dataAvailOrg,CMT} called a  \emph{stalling attack}, where a side blockchain node commits the hash of a block to the trusted blockchain but makes the block itself unavailable to other side blockchain nodes. Authors in \cite{AceD} proposed a scalable solution to the above attack by introducing a data availability oracle between the trusted blockchain and side blockchains. The oracle consists of nodes whose goal is to collectively ensure that the block is available, even if some of the oracles nodes are malicious. Nodes in the oracle layer accept the block from a side blockchain node (who wishes to commit its hash to the trusted blockchain), and push the hash commitment only if the block is available to the system. The goal is to share (\emph{disperse}) the block among the oracle nodes in a storage and communication efficient way to check the block availability. 
The solution in \cite{AceD} involves using a Low-Density Parity-Check (LDPC) code to generate coded chunks from the block such that each oracle node receives different coded chunks, and \debbb{using \emph{incorrect-coding proofs} \cite{dataAvailOrg,CMT} to ensure}  that the block is correctly coded. 
A dispersal protocol ensures that the oracle nodes receive sufficient coded chunks that guarantee that the original block can always be decoded by a peeling decoder using the coded chunks sent to the oracle nodes, (i.e., the block is available), 
even in the presence of malicious oracle nodes.

\vspace{-0.05cm}
Stopping sets are a set of variable nodes (VNs) of an LDPC

\noindent
code that if erased \deb{prevent} a peeling decoder from decoding the block. Formally, a set of VNs of an LDPC code is called a stopping set if every check node (CN) connected to this set of VNs is connected to it at least twice. 
To guarantee block availability in the presence of malicious oracle nodes, the dispersal protocol defined in \cite{AceD} requires every subset of oracle nodes of a particular size to receive at least $M - M_{\min} +1$ distinct coded chunks, where $M$ and $M_{\min}$ are the blocklength and minimum stopping set size of the LDPC code, respectively. As a result, the communication cost associated with the dispersal is inversely proportional to the minimum stopping set size of the LDPC code. 
\deb{Thus, authors in \cite{AceD} focused on LDPC code constructions with large minimum stopping set size for their dispersal protocol. This combination of dispersal protocol and LDPC construction may not necessarily be optimal in terms of communication costs.}
In this paper, we design a new dispersal protocol that considers the multiplicity of small stopping sets and provide a specialized LDPC code construction based on the Progressive Edge Growth (PEG) algorithm \cite{PEG}, which we call the \emph{dispersal-efficient} PEG (DE-PEG) algorithm, that aims at minimizing the communication cost within our protocol. We demonstrate a significantly lower communication cost using our specialized LDPC construction and dispersal protocol in comparison to \cite{AceD}.
\debbb{Our techniques support a wider range of system parameters allowing for more flexibility in system design such as scaling the number of oracle nodes while still allowing scalability of the block size as \cite{AceD} thereby providing a much more scalable solution to the stalling attack problem.}
\debbb{Previously, channel coding has been extensively used to mitigate issues such as data availability, storage, and communication in blockchain systems \cite{dataAvailOrg,CMT,SSskewarxiv,cover,polyshard,SeF,patternederasure,secregenerating,crossshard,networkcodedPBFT,erasurelowstorage,secureraptor}.}

The rest of this paper is organized as follows. In Section \ref{section:preliminaries}, we describe the preliminaries and system model. In Section \ref{sec:design_idea}, we provide our new dispersal protocol and motivate our LDPC design criterion. The DE-PEG algorithm is described in Section \ref{sec:DE-PEG}. \deb{Finally, the simulation results and concluding remarks are presented in Section \ref{sec:simulation_results}}.

\vspace{-0.25cm}
\section{Preliminaries and System Model}\label{section:preliminaries}
\begin{figure}[t]
    \centering
    \includegraphics[scale=0.41]{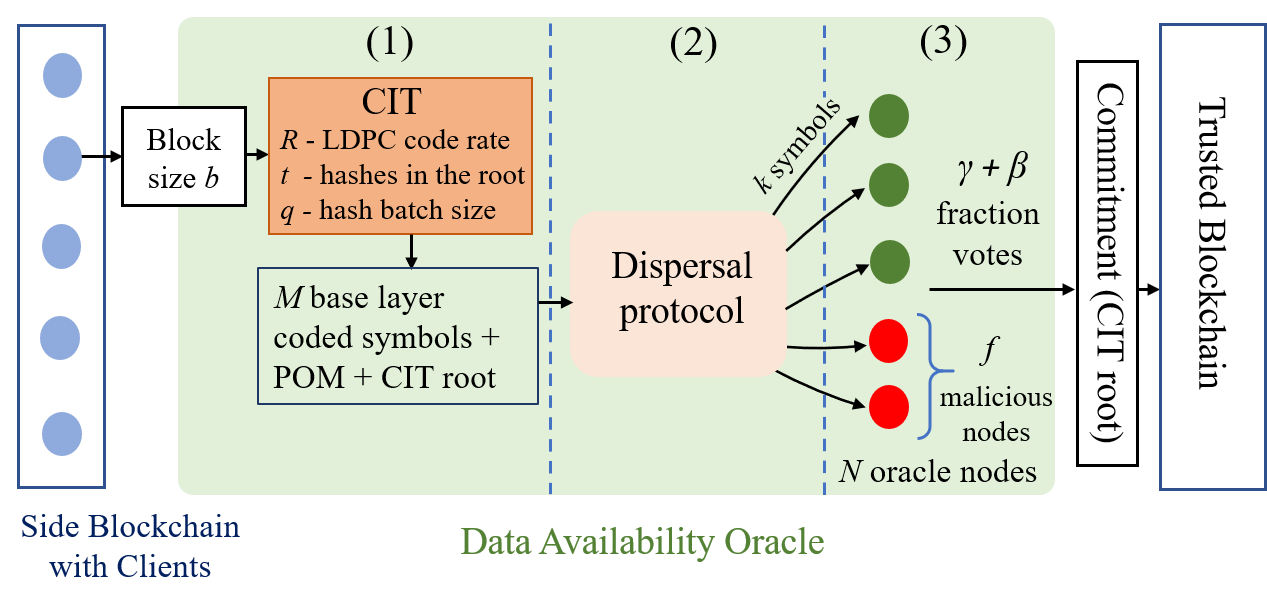}
    \vspace{-0.65cm}
  \caption{System Model. The network consists of oracle nodes and side blockchain nodes (called \emph{clients}), where clients propose blocks to the oracle nodes to commit to the trusted blockchain. Oracle nodes verify the correctness of each received block and submit the block commitment to the trusted blockchain if the block is available.}
    \vspace{-0.2cm}
    \label{fig:oracle_model}
    \vspace{-0.35cm}
\end{figure}

\vspace{-0.2cm}
In this paper, we assume the blockchain and data availability oracle model of \cite{AceD} \deb{and is summarized in Fig. \ref{fig:oracle_model}}. Suppose that there are $N$ oracle nodes and an adversary is able to corrupt a fraction $\beta$ of them, where $\beta < \frac{1}{2}$, such that the maximum number of malicious oracles nodes is $f = \left \lceil{\beta N}\right \rceil$.  
When a client proposes a block of size $b$, it first generates a special Coded Merkle Tree \cite{CMT}, called a Coded Interleaving Tree (CIT) introduced in \cite{AceD}, with the data chunks of the block as leaf nodes of the CIT. A CIT is a coded version of a regular Merkle tree \cite{Bitcoin} and is constructed by applying a rate-$\tr$ systematic LDPC code to each layer of the Merkle tree before hashing the layer to generate its parent layer. Details regarding the CIT construction can be found in \cite{AceD} and in 
Appendix \ref{appendix:CIT_construction}.
In particular, the CIT has $M$ base layer coded chunks (symbols),
each with an associated Proof of Membership (POM), which consists of a systematic (data) symbol and a parity symbol from each CIT layer. The CIT has a root with $t$ hashes and in each layer $q$ hashes are batched together into a data chunk for the layer. 
The data availability oracle functions in the following way as shown in Fig. \ref{fig:oracle_model}:
\vspace{-0.1cm}
\begin{enumerate}[wide, labelwidth=!, labelindent=0pt]
    \item When a client proposes a block of size $b$, it constructs its CIT, which generates a set of base layer coded symbols, each with an associated POM, and a CIT root. 
    
    \item The client then uses a dispersal protocol to disperse the base layer coded symbols,
    their associated POMs, and the CIT root to the $N$ oracle nodes. The dispersal protocol specifies the base layer coded chunks each oracle node should receive, each receiving $k$ of them (with their POMs) and the CIT root. 
    
    \item Each of the oracle nodes, on receiving the specified $k$ coded chunks check their correctness (i.e., whether they satisfy the associated POM with the root). The dispersal is accepted if $\gamma + \beta$ fraction of the nodes vote that they individually received all correct coded chunks, for a parameter $\gamma \leq 1 - 2\beta$ defined in the dispersal protocol. In this case, the CIT root is committed to the trusted blockchain and each of the oracle nodes store the $k$ coded chunks they received to allow for future block retrieval. The CIT prevents clients from performing incorrect coding of the block  via an incorrect-coding proof \cite{AceD}.
\end{enumerate}

\vspace{-0.1cm}
The focus of this paper is to design a dispersal protocol and an associated LDPC code to reduce the communication cost of the dispersal process. The dispersal protocol must satisfy the \emph{availability} condition: whenever the root of the CIT is committed to the trusted blockchain, an honest client must be able to decode each CIT layer using a peeling decoder by requesting for the coded chunks stored at the oracle nodes. 
 Each CIT layer in \cite{AceD} is constructed using random LDPC codes that with high probability have a stopping ratio (minimum stopping set size divided by the blocklength) $\alpha^*$. The dispersal protocol in \cite{AceD} is designed such that every $\gamma$ fraction of the oracle nodes receive more than $1-\alpha^*$ fraction of distinct base layer coded chunks.
  Moreover,  it was shown in \cite{AceD} that the POMs of any $\eta$ fraction of distinct base coded chunks have at least $\eta$ fraction of distinct coded chunks from each CIT layer (we call this the \emph{repetition property}). Thus, the dispersal protocol ensures that every $\gamma$ fraction of nodes also have more than $1-\alpha^*$ fraction of distinct coded chunks from each CIT layer.
Hence, when a root is committed, due to 3), there is a $\gamma$ fraction of honest oracle nodes who have more than $1-\alpha^*$ fraction of coded symbols from each CIT layer, allowing a peeling decoder to decode each layer ensuring \emph{availability}. 

  Let the CIT have $l$ layers and $n_j$ coded chunks in layer $j$, $1 \leq j \leq l$, where $n_l = M$. Let $H_j$ denote the parity check matrix of the LPDC code used in layer $j$ which has $n_j$ columns $\{v^j_1, v^j_2, \ldots, v^j_{n_j}\}$ %
  (we drop the superscript $j$ based on context). Let $\tG_j$ denote the Tanner graph (TG) representation of $H_j$, where we also refer to $v_i$ as the $i^{th}$ VN in $\tG_j$ and rows of $H_j$ as CNs in $\tG$. 
  A cycle of length $g$ is called a $g$-cycle. 
  For a set $T$, let $\vert T \vert$ denote its cardinality. Let the dispersal protocol be defined by the set $\tC = \{A_1, A_2, \ldots, A_{N}\}$, where $A_i$ denotes the set of base layer coded chunks sent to oracle node $i$ and $\vert A_i \vert = k$. For a set $S$ of VNs, let $\tnei(S)$
 be the set of oracle nodes who have at least one coded chunk corresponding to the VNs of $S$. %
 Let the hashes of each coded block be of size $y$. Let $H_e(p) = -p\ln(p) - (1-p)\ln(1-p)$.

\vspace{-0.1cm}
\begin{definition}\label{defn:valid}
 Protocol $\tC$ is called $\eta$-valid for layer $j$ if every $\gamma$ fraction of oracle nodes have $>$ $\eta$ fraction of distinct layer $j$ coded chunks. Similarly, $\tC$ is $\mu$-SS-valid for layer $j$ if every $\gamma$ fraction of oracle nodes have $> n_j - \mu$ distinct layer $j$ coded chunks.
If no layer is specified, we refer to the base layer. %
\end{definition}
\vspace{-0.1cm}
\noindent
Note that a protocol that is $\mu$-SS-valid for layer $j$ is also $\left(\frac{n_j - \mu}{n_j}\right)$-valid for layer $j$. 
Due to the \emph{repetition} property, if the base layer \debbb{is} $\mu$-SS-valid, we can determine $\tilde{\mu}$ such that the  protocol is $\tilde{\mu}$-SS-valid for layer $j$. Thus, we majorly talk about the base layer and drop the specification of the layer according to Definition \ref{defn:valid}. In \cite{AceD}, the dispersal protocol is required to be $(1 - \alpha^*)$-valid for all layers.
A protocol which is $\mu$-SS-valid for layer $j$ can guarantee that a client will be able to decode layer $j$ of the CIT using a peeling decoder when the block is committed and stopping sets of size $ < \mu$ do not exist in $H_j$.

In \cite{AceD}, elements of $A_i$ are randomly chosen with replacement from the set of $M$ base layer coded chunks. For such a design, 
it was shown in \cite{AceD} that for $ k > \frac{M}{N\gamma}\ln\frac{1}{1-\eta}$,     $\mathrm{Prob}(\tC \text{ is not } \eta\text{-valid}) \leq \exp(N H_e(\gamma)- M f(\eta, \rho))  := P^{\mathrm{UB}}(\eta,N,M,k,\gamma)$, 
where 
$\rho = \frac{\gamma N k}{M}$ and $f(\eta, \rho) = \frac{(e^{\rho}(1-\eta)-1)^2}{e^{\rho}(e^{\rho}(1-\eta)+1)}$ is a positive function.
It is clear that $M$ can be made sufficiently large to make $P^{\mathrm{UB}}(\eta,N,M,k,\gamma)$ arbitrarily small. This principle was used in \cite{AceD} to randomly design the dispersal protocol. However, as we show next, to make $P^{\mathrm{UB}}(\eta,N,M,k,\gamma)$ smaller than a given threshold probability $p_{th}$, for a fixed $M$,  there is a limit on the number of oracle nodes the system can support. 

\vspace{-0.15cm}
\begin{lemma}\label{lemma:N_upper_bound}
Let $N^{UB} = \frac{M(1-\eta) + \ln(p_{th})}{H_e(\gamma)}$ and $\Bar{\eta} = 1 - \eta$.
If $N \geq N^{UB}$,
$P^{\mathrm{UB}}(\eta,N,M,k,\gamma) > p_{th}$ $\forall k > \frac{M}{N\gamma}\ln\frac{1}{\Bar{\eta}}$.  If $N < N^{UB}$, then $P^{\mathrm{UB}}(\eta,N,M,k,\gamma) \leq p_{th}$ for $k \geq k^f_{\min} :=\frac{M}{N\gamma}\ln\left(\frac{-(2\Bar{\eta}+v) - \sqrt{8\Bar{\eta}v + v^2}}{2\Bar{\eta}(v-\Bar{\eta})}\right)$, where $v = \frac{NH_e(\gamma) - \ln(p_{th})}{M}$. 
\end{lemma}
\vspace{-0.25cm}
\begin{proof}
\debbb{The proof relies on algebraic manipulation of $P^{\mathrm{UB}}()$ and can be found in Appendix \ref{appendix:proof_N_upper_bound}.}
\end{proof}
\vspace{-0.15cm}
Thus, the dispersal protocol used in \cite{AceD} cannot guarantee with high probability to be $(1-\alpha^*)$-valid for all $(N,M)$ pairs. This feature is undesirable and we would like for a given $M$, any number of oracle nodes to be supported by the protocol. The problem is alleviated if each $A_i$ gets $k$ distinct coded chunks chosen uniformly at random from the $M$ base layer coded chunks which we consider in our design idea.

\vspace{-0.12cm}
\section{Design Idea: Secure Stopping Set Dispersal}\label{sec:design_idea}
\vspace{-0.05cm}
\begin{definition}
 A protocol $\tC = \{A_1, \ldots, A_{N}\}$ is called a $k$-dispersal if each $A_i$ is a $k$ element subset chosen uniformly at random with replacement from all the $k$ element subsets of the M base layer coded chunks.
\end{definition}
\vspace{-0.15cm}
We analyze the minimum number of distinct coded chunks $k$ to disperse to each oracle node so that the protocol $\tC$ is $\mu$-SS-valid with probability at least $1 -p_{th}$. 

\vspace{-0.12cm}
\begin{lemma}\label{lemma:coupon_group}
For a  $k$-dispersal protocol, for the base layer,
\vspace{-0.1cm}
$$\mathrm{Prob}(\tC \text{ is not } \mu\text{-SS-valid}) \leq e^{NH_e(\gamma)}P_{f},$$ \vspace{-0.8cm}
\begin{align*}
    \text{where } P_{f} =  \text{\hspace{-0.05cm}}\sum_{j=0}^{M-\mu} (-1)^{M-\mu -j} {M \choose j} {M - j - 1 \choose \mu -1}\left[\frac{{j \choose k}}{{M \choose k}}\right]^{\gamma N}
\end{align*}
\end{lemma}\vspace{-0.45cm}
\begin{proof}
\debbb{The proof utilizes the fact that the process of a given $\gamma N$ nodes sampling with replacement the $k$ element subsets of the $M$ base layer chunks is known as the coupon collector's problem with group drawings \cite{CouponGroup}. The proof is provided in Appendix \ref{appendix:coupon_group}.}
\end{proof}
\vspace{-0.17cm}
Now, $e^{NH_e(\gamma)}P_{f}$ can be made smaller than an arbitrary threshold $p_{th}$ by choosing a sufficiently large $k$. Let $k^*(\mu,N,M,\gamma,p_{th})$ be the smallest $k$ such that $e^{NH_e(\gamma)}P_{f} \leq  p_{th}$. Thus, a $k^*(\mu,N,M,\gamma,p_{th})$-dispersal will be $\mu$-SS-valid for the base layer with probability $\geq 1 - p_{th}$. The associated communication cost is $N Xk^*(\mu,N,M,\gamma)$, where $X$ is the total size of one base layer coded chunk along with its POM. To ensure availability, if we set $\mu = M_{\min}$, we see that the communication cost is directly affected by the minimum stopping set size. Thus, for a $k^*(M_{\min},N,M,\gamma, p_{th})$-dispersal protocol, the best code design strategy is to design LDPC codes with large minimum stopping set sizes which is considered to be a hard problem\debb{\cite{SSelim,SSelimSurvey}}. In this work, we modify the above dispersal protocol to reduce the communication cost. We then provide a specialized LDPC code construction aimed at minimizing the communication cost associated with the modified dispersal protocol. 

\vspace{-0.12cm}
\begin{definition}
 A stopping set $S$ is said to be securely dispersed by a dispersal protocol $\tC$ if $\vert \tnei(S) \vert \geq f+1$.
\end{definition}
\vspace{-0.1cm}
Since $f$ is the maximum number of malicious oracle nodes, for a stopping set that is securely dispersed, at least one honest oracle node will have a coded chunk corresponding to a VN of $S$ and hence the peeling decoder will not fail due to $S$ and can continue the decoding process. Based on this principle, we consider the following dispersal protocol:

\vspace{-0.15cm}
\begin{strategy}\label{dispersal_protocol} ($k^*$-secure dispersal)
For  $\mu \geq M_{\min}$,

\noindent
let $\tS_j$ be the set of stopping sets of layer $j$ (i.e., of $H_j$) of size less than $n_j - \left \lceil{\left(\frac{M - \mu+1}{M}\right)n_j}\right \rceil +1$. Our dispersal protocol consists of two dispersal phases. In the first phase (called the secure phase), all stopping sets in $\tS_j$, $1 \leq j \leq l$, are securely dispersed. This is followed by a $k^*(\mu,N,M,\gamma,p_{th})$-dispersal protocol (called the valid phase).
\end{strategy}

\vspace{-0.35cm}
\begin{lemma}\label{lemma:dispersal_availability}
Dispersal protocol \ref{dispersal_protocol} guarantees availability with probability $\geq 1 - p_{th}$.
\end{lemma}
\vspace{-0.45cm}
\begin{proof} Proof is provided in 
 Appendix \ref{appendix:dispersal_availability}.
For $d_j = (n_j - \left \lceil{\left(\frac{M - \mu+1}{M}\right)n_j}\right \rceil + 1)$, we use the \emph{repetition} property to show that with probability $\geq  1 - p_{th}$, the protocol is $d_j$-SS-valid for each layer $j$, and all stopping sets of sizes $< d_j$ are securely dispersed. These conditions guarantee availability. 
\end{proof}

\vspace{-0.15cm}
We use the following greedy procedure to securely disperse all stopping sets in $\tS_j$, $1 \leq j \leq l$.
Let $\tCg_j$ be a set of VNs of $H_j$ with the property that for all $S \in \tS_j$, $\exists$ $v \in \tCg_j$ such that $v$ is part of $S$. Note that if each VN in $\tCg_j$ is sent to $(f+1)$ oracle nodes, all $S \in \tS_j$ will be securely dispersed. We obtain $\tCg_j$ in the following greedy manner: Initialize $\tCg_j = \emptyset$.  Find a VN $v$ that is part of the maximum number of stopping sets in $\tS_j$, add the VN to $\tCg_j$ and remove all stopping sets in $\tS_j$ that have $v$. We repeat the process until $\tS_j$ is empty.
Let the VNs in each set $\tCg_j$ be ordered according to the order they were added to $\tCg_j$. For each $j$, we permute the columns of $H_j$ such that the VNs in $\tCg_j$ appear as columns $1, 2, \ldots, \vert \tCg_j \vert$, the rest of the columns are randomly ordered. Note that the $H_j$'s after the column permutation are used to build the CIT.
Now, our secure phase is designed as follows: the design starts from layer $l$ and moves iteratively up the tree till layer 1. For each layer $j$, if all VNs corresponding to the first $\vert \tCg_{j} \vert$ columns of $H_{j}$ are marked as \emph{dispersed}, we mark layer $j$ as \emph{complete} and move to layer $j-1$, else,  we disperse the remaining coded chunks corresponding to the first $\vert \tCg_{j} \vert$ columns of $H_{j}$ that are not marked \emph{dispersed} by randomly selecting $f+1$ oracle nodes to send each of the coded chunk with its POMs. For each layer $i$ above layer $j$, coded chunks that were sent to $(f+1)$ nodes as part of POMs of the coded chunks of layer $j$ in the previous step are marked as \emph{dispersed}. We mark layer $j$ as \emph{complete} and \vspace{0.08cm}proceed to layer $j-1$. We continue until layer 1 is \emph{complete}. 
\hspace*{0.27cm}Note that by initially permuting the columns of $H_j$'s, we have ensured that when a coded chunk of a particular layer $j$ with its POMs  are sent to $(f+1)$ nodes, the systematic symbol of the POMs from each layer $i$ above layer $j$ are exactly the VNs in the first $\vert \tCg_i \vert$ columns of $H_i$ that we require to send to $(f+1)$ nodes to securely disperse $\tS_i$. If the POM for layer $i$ is outside the first $\vert \tCg_i \vert$ columns (happens if $\vert  \tCg_j\vert > \vert \tCg_i\vert$), this would imply that layer $i$ is already \emph{complete}.

Let $X_j$ be the size of one coded chunk of layer $j$ along with its POMs which involve a data and parity symbol from each layer above layer $j$. Also, let $t_j = \max_{i \in \{j+1, \ldots, l\}} \vert \tCg_i \vert$. As such, $X_l = \frac{b}{\tr M} + y(2q-1)(l-1)$ and $X_j = qy + y(2q-1)(j-1)$, $1\leq j < l$, \cite{AceD} (details can also be found in 
Appendix \ref{appendix:CIT_construction}.
The total communication cost $\mathrm{C}^T$ for Dispersal Protocol \ref{dispersal_protocol} is $\mathrm{C}^T = Nty + \mathrm{C}^s +\mathrm{C}^v$, where $\mathrm{C}^s$ and $\mathrm{C}^v$ are the costs associated with the secure and valid phases, respectively, and $Nty$ is the cost of dispersing the CIT root. Now, $\mathrm{C}^v = Nk^*(\mu,N,M,\gamma, p_{th})X_l$ and $\mathrm{C}^s$ can be calculated as
$\mathrm{C}^{s} =  (f+1)\left[ \vert \tCg_l \vert X_l  + \sum_{j=1}^{l-1}\max\left((\vert \tCg_j \vert - t_j),0\right)X_j\right],$
where we have made the assumption that each $\vert \tCg_j \vert$ is smaller than $\tr n_j$ which is true for small $\mu$ and since $\tr n_j$ is the total number of systematic variable nodes. The communication cost of the secure phase depends strongly on $\vert \tCg_l \vert$ as the base layer involves data chunks whose sizes are larger than the chunks of the higher layers which are concatenations of hashes. Thus, we can reduce the total cost by designing LDPC codes that have small $\vert \tCg_l \vert$. We provide the construction in the next section.

\vspace{-0.1cm}
\section{Dispersal-Efficient PEG Algorithm}\label{sec:DE-PEG}
\vspace{-0.1cm}
Algorithm \ref{alg:DE-PEG} presents our DE-PEG algorithm that constructs a TG $\tTG$ with $M$ VNs, $J$ CNs, and VN degree $d_v$ that results in a small size of $\tCg_{l}$. Note that the same algorithm is used for all layers to reduce the sizes of $\tCg_{j}$. Since stopping sets in LDPC codes are made up of cycles \cite{WesselSS}, the DE-PEG algorithm focuses on cycles as they are easier to optimize. In the algorithm, all ties are broken  randomly.

\vspace{-0.2cm}
\begin{algorithm}
\caption{DE-PEG Algorithm}\label{alg:DE-PEG}
\begin{algorithmic}[1]
\State \textbf{Inputs:} $M$, $J$, $d_v$, $g_{\max}$, $T_{th}$ \textbf{Output:} $\tTG$
\State \textbf{Initialize} $\tTG$ to $M$ VNs, $J$ CNs and no edges, $\tL = \emptyset$
\For{$j=1$ to $M$}
    \For{$e=1$ to $d_{v}$}
    \State [$\tK , g$] = \tPEG$(\tTG,v_j)$
        \If{$g > g_{\max}$} 
    $\tcn^{s}$ = uniformly random CN in $\tK$ 
        \Else\Comment{\textit{($g$-cycles, $g \leq g_{\max}$, are created)}}
        \For{each CN $c$ in $\tK$}
        \State $\tL_{cycles}$ = $g$-cycles formed due to $c$
        \State $\Bar{s}[c] = $ \tgreedy($\tL \cup \tL_{cycles},v_j$)
        
        \EndFor
        \State
            $\tcn^{s}$ = CN in $\tK$ with minimum $\Bar{s}[c]$
        \State
        $\tL^s$ = $g$-cycles formed due to $\tcn^{s}$ with EMD $\leq$ $T_{th}$
        \State
        $\tL = \tL \cup \tL^s$
        \EndIf
    \State  $\tTG = \tTG \cup \mathrm{edge}\{\tcn^{s},v_j\}$

    \EndFor
\EndFor
\end{algorithmic}
\end{algorithm}
\vspace{-0.2cm}
 The algorithm uses the concept of the extrinsic message degree (EMD) of a set of VNs which is the number of CN neighbours singly connected to the set of VNs \cite{EMD} and is calculated using the method described in \cite{EMDcalculation}. EMD of a cycle is the EMD of the VNs involved in the cycle. Cycles with low EMD are more likely to form a stopping set and we consider them as \emph{bad} cycles. 
The algorithm also uses a procedure \tgreedy($\tilde{\tL},v$) which takes as input a list $\tilde{\tL}$ of cycles, and outputs $\vert \Bar{S} \vert $, where $\Bar{S}$ is a set of VNs with the property that for every cycle $C$ in $\tilde{\tL}$, $\exists$ a VN in $\Bar{S}$ that is part of $C$, and $\Bar{S}$ is obtained in a manner similar to that of obtaining $\tCg_j$ from $\tS_j$ described in Section \ref{sec:design_idea}, however, by ignoring the VN $v$ during the greedy selection procedure.

The \debbb{PEG algorithm \cite{PEG}} builds a TG by iterating over the set of VNs and for each VN $v_j$, establishing $d_v$ edges to it. 
For establishing the $e^{th}$ edge, there are two situations that the algorithm encounters: i) addition of an edge is possible without creating cycles; ii) addition of an edge creates cycles.
In both the situations, the PEG algorithm finds a set of \emph{candidate} CNs to connect $v_j$ to, that maximises the girth of the cycles formed. We do not go into the detailed procedure followed by \cite{PEG} to find the set of candidate CNs, but assume a procedure {\fontfamily{qcr}\selectfont PEG}$(\mathcal{G},v_j)$ that provides us with the set of candidate CNs $\tK$ for establishing a new edge to VN $v_j$ under the TG setting $\mathcal{G}$ according to the \debbb{PEG algorithm}. We assume that the set $\tK$ only contains CNs with the minimum degree under the TG setting $\mathcal{G}$. For situation ii), the procedure returns the cycle length $g$ of the smallest cycles formed when an edge is established between any CN in $\tK$ and $v_j$. For situation i) $g = \infty$ is returned. When $g > g_{\max}$ is returned,
we follow the \debbb{PEG algorithm} and select a CN randomly from $\tK$. 

During the course of the DE-PEG algorithm, we maintain a list $\tL$ of \emph{bad} cycles of lengths $\leq g_{\max}$ that had EMD less than or equal to some threshold $T_{th}$ when they were formed. 
In the algorithm, when cycle length $g \leq g_{\max}$ is returned by the \tPEG$()$ procedure, for each CN $c \in \tK$, $g$-cycles are formed when an edge is added between $c$ and $v_j$. These cycles are listed in $\tL_{cycles}$ (line 9). We use \tgreedy($\tL \cup \tL_{cycles}, v_j$) to get $\Bar{s}[c]$ (line 10), for each CN $c$ in $\tK$. Our CN selection procedure is to select a CN from $\tK$ that has the minimum $\Bar{s}$  (line 10).
Once this CN is selected, we update the list of \emph{bad} cycles as follows: of all the $g$-cycles formed due to the addition of an edge between $\tcn^s$ and $v_j$, we find the list of $g$-cycles $\tL^s$ that have EMD $\leq T_{th}$ (line 12) and add them to $\tL$ (line 13). We then update the TG $\tTG$ (line 14). 
 \debbb{The intuition is} that since we want the stopping sets in $\tS_l$ to produce a small $\tCg_l$ by a greedy procedure, we select CNs such that a similar greedy procedure produces small $\vert\Bar{S}\vert$ on the bad cycles which are more likely to form stopping sets. Note that a similar PEG algorithm was provided in \cite{SSskewarxiv} but it had a different design objective compared to this paper.
 
 \vspace{-0.15cm}
 \begin{remark}
 In the DE-PEG algorithm, \emph{\tgreedy}($\tL \cup \tL_{cycles}, v_j$) ignores the VN $v_j$ while forming the greedy set of VNs to find $\vert\Bar{S}\vert$ as $v_j$ is part of all the cycles formed by all CNs $c$ in $\tK$ and ignoring $v_j$ allows to better distinguish between the CNs in terms of set sizes $\Bar{s}$. 
 While the DE-PEG algorithm is based on cycles, Dispersal Protocol \ref{dispersal_protocol} uses stopping sets $\tS_j$ to find $\tCg_j$ for the secure phase. %
 \end{remark}

 \vspace{-0.4cm}
 \section{\deb{Simulations and Conclusion}}\label{sec:simulation_results}
 \vspace{-0.15cm}
 In this section, we present the performance of the codes designed using the DE-PEG algorithm when using the $k^*$-secure dispersal protocol. 
 To demonstrate the benefits, we consider a baseline system that uses codes constructed using the original PEG algorithm and uses $k$-dispersal with $k$ chosen such that for all layers $j$, $1\leq j \leq l$, the $k$-dispersal is $M^j_{\min}$-SS-valid, where $M^j_{\min}$ is the minimum stopping set size of layer $j$ (and $M_{\min} = M^l_{\min}$).
 To compute a lower bound on the total communication cost using Dispersal Protocol \ref{dispersal_protocol}, we consider a code that has $\tS_j = \emptyset$, \deb{$1\leq j \leq l$}, i.e., for the given $\mu$, has costs only due to $k^*(\mu,N,M,\gamma,p_{th})$-dispersal and the root, and no cost due to the secure phase.  This is equivalent to designing codes having larger minimum stopping set sizes which is considered hard. 
 We use the following parameters for simulations: $b = 1$MB, $y = 32$ Bytes, $t = 32$, $q  =4$, $l = 4$, $M = 256$, $\tr = 0.5$,  $\gamma = 1 - 2\beta$, $p_{th} = 10^{-8}$ (specified if otherwise). The CIT thus has 4 layers with $n_4 = M = 256$, $n_3 = 128$, $n_2 = 64$ and $n_1 = 32$. For the LDPC codes constructed using the DE-PEG algorithm, we use $g_{\max} = 8$ for layer 3 and 4 and $g_{\max} = 6$ for layer 1 and 2, $d_v = 4$, and $T_{th} = 5$ (provides the best results from a range of thresholds tested). For the base layer, the PEG and DE-PEG codes constructed have $M_{\min} = 17$ and $18$ respectively. All communication costs are in GB. Costs $\mathrm{C}^s$, $\mathrm{C}^v$ and $\mathrm{C}^T$ are calculated using equations described in Section \ref{sec:design_idea}. 
 
 Table \ref{table:mu_variation} compares the communication cost achieved by the PEG and DE-PEG algorithm with the $k^*$-secure dispersal protocol as the value of $\mu$ is varied.
 We see that as $\mu$ is increased, the value of $k^*$ decreases and $\mathrm{C}^v$ decreases. The table next shows the 4-tuple $(|\tCg_1|,|\tCg_2|, |\tCg_3|, |\tCg_4|)$ for the PEG and DE-PEG algorithm and we see that the DE-PEG algorithm always results in lower values thus resulting in a lower cost $\mathrm{C}^s$ during the secure phase compared to the PEG algorithm. Note that, as $\mu$ is increased, $\mathrm{C}^s$ increases. Finally, we look at the total cost $\mathrm{C}^{T}$, which is lowest for $\mu = 20$ for both the PEG and DE-PEG algorithms, and are 0.425GB and 0.528GB lower, respectively, compared to the baseline ($\mathrm{C}^{T}$ = 5.125GB) at  $\mu  = 17$. Interestingly, $\mathrm{C}^{T}$ does not monotonically decrease with $\mu$. Note that the lower bound on $\mathrm{C}^{T}$ at $\mu = 20$ is 0.687GB lower than the baseline.

\setlength{\tabcolsep}{3pt}
\begin{table}
\caption{Communication costs achieved by  $k^*$-secure dispersal for various choices of $\mu$ using the PEG and DE-PEG algorithm for $N = 9000$, $\beta = 0.49$. Lower bound on $\mathrm{C}^T$ for $\mu = 20$ is 4.438GB.}\label{table:mu_variation}
 \vspace{-5pt}
\centering
\resizebox{!}{1.15cm}{
\begin{tabular}{| c | c | c | c | c | c |c | c| c|}
\hline
 \multirow{2}{*}{$\mu$} & \multirow{2}{*}{$k^*$} & 
\multirow{2}{*}{$\mathrm{C}^{v}$} &
\multicolumn{2}{|c|}{$(|\tCg_1|,|\tCg_2|, |\tCg_3|, |\tCg_4|)$}&
\multicolumn{2}{|c|}{$\mathrm{C}^{s}$} &
\multicolumn{2}{|c|}{$\mathrm{C}^{T}$}\\
& & & PEG & DE-PEG & PEG & DE-PEG & PEG & DE-PEG \\
\hline
17 & 67 & 5.116 & (0,0,0,0) & (0,0,0,0) & 0 & 0 & 5.125 & 5.125\\
18	&64	&4.887	&(0,0,0,1)	&(0,0,0,0)&	0.037	&0	&4.933&	4.896 \\
19	&61	&4.658	&(0,0,1,3)	&(0,0,0,1)	&0.112	&0.037	&4.779	&4.704\\
20	&58	&4.428	&(0,0,1,7)	&(0,0,0,4)	&0.262	&0.149	&4.700	&4.587\\
21 &56	&4.276	&(0,1,2,14)	& (0,1,0,13)	&0.524	& 0.486	&4.809 & 4.771\\
\hline
\end{tabular}}
\vspace{-0.2cm}
\end{table}
 
\setlength{\tabcolsep}{1pt} 
\begin{table}
\caption{Comparison of total communication cost of our work with \cite{AceD} for $\beta = 0.49$, $M = 256$. For each $p_{th}$, $N$ is the maximum no. of oracle nodes permissible for the oracle of \cite{AceD} and $k_{\min}$ is the minimum number of chunks at each oracle node for $\eta = 1 - \alpha^*$ and are computed using Lemma \ref{lemma:N_upper_bound} (Note that $k_{\min} = \left\lceil{k^f_{\min}}\right\rceil$). 
}
\label{table:comparison}
\vspace{-5pt}
\centering
\begin{tabular}{|c|c|c c  c |c c |c c  c |}
\hline
&&
\multicolumn{3}{|c|}{Oracle \cite{AceD}}&
\multicolumn{5}{|c|}{Our Work}\\
\hline
 \multirow{2}{*}{$p_{th}$} & \multirow{2}{*}{$N$}&
 \multicolumn{3}{|c|}{$\alpha^* = 0.125$}&
 \multicolumn{2}{|c|}{$\mu =17$} & \multicolumn{3}{|c|}{$\mu = 20$}\\
  &&$k_{\min}$ & $\mathrm{C}^T_{full}$ & $\mathrm{C}^T_{distinct}$ & $k^*$ & $\mathrm{C}^T_{baseline}$ & $k^*$ & $\mathrm{C}^T_{PEG}$ & $\mathrm{C}^T_{DE-PEG}$\\
 \hline
 $10^{-8}$& 138 &895 & 1.048  & 0.2909 & 207 & 0.2425 & 199 & 0.2372 & 0.2355\\
 $10^{-6}$ &185 & 671 & 1.053 & 0.3729 & 184 & 0.2890 & 175 & 0.2803 & 0.2780\\
 $10^{-4}$ & 232 & 539 & 1.061 & 0.4430 & 164 & 0.3231 & 155 & 0.3122 & 0.3092\\
\hline

\end{tabular}
\vspace{-0.4cm}
\end{table}

 \begin{figure}[t]
    \centering
\begin{minipage}{0.99\linewidth}
\vspace{-0.25cm}
\begin{tikzpicture}
  \node (img)
  {\includegraphics[scale=0.27]{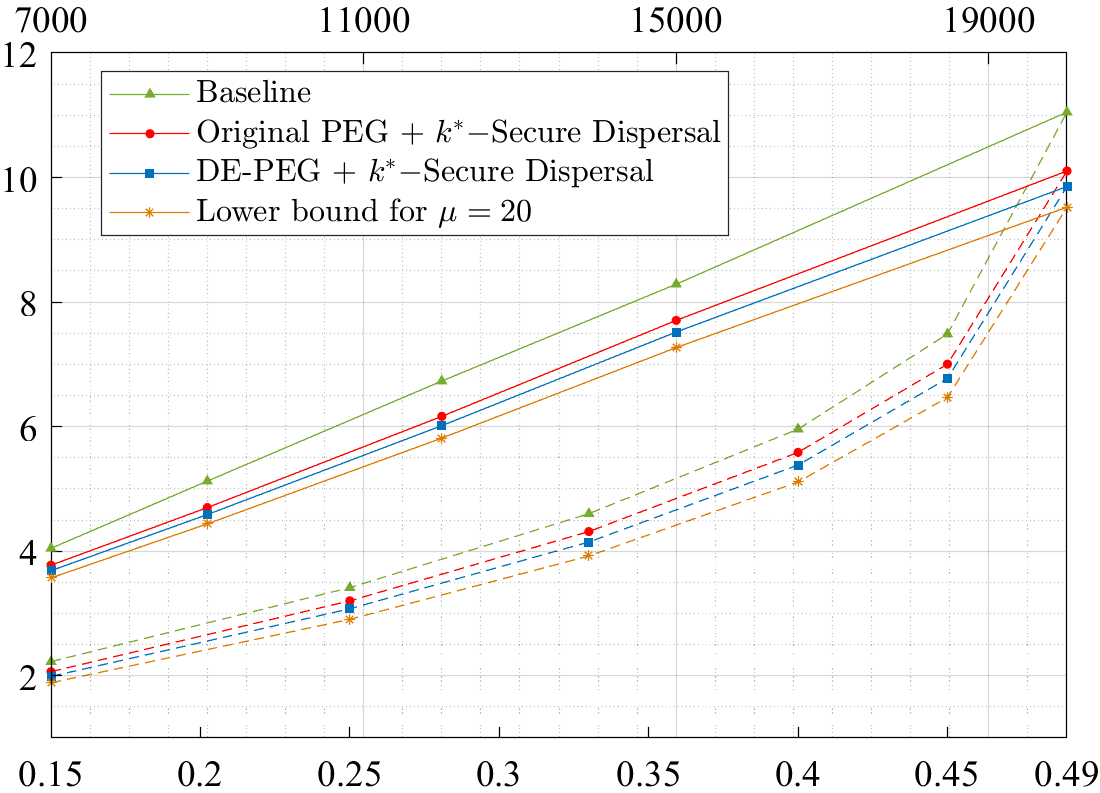}};
  \node[below=of img, node distance=0cm, yshift=1.2cm,font=\color{black}] {$\beta$ (- - - -)};
  \node[above=of img, node distance=0cm, yshift=-1.2cm,font=\color{black}] {$N$ (------)};
  \node[left=of img, node distance=0cm, rotate=90, anchor=center,yshift=-1cm,font=\color{black}] {$\mathrm{C}^T$ (in GB)};
 \end{tikzpicture}
 \end{minipage}
 \vspace{-7pt}
\caption{$\mathrm{C}^T$ for various coding schemes and dispersal protocols vs. (a)  $N$ at $\beta = 0.49$ (solid plots, top axis) and (b) $\beta$ at $N = 20000$ (dotted plots, bottom axis). For $k^*$-secure dispersal $\mu = 20$ is used.}
\label{fig:combined}
     \vspace{-0.5cm}
\end{figure}
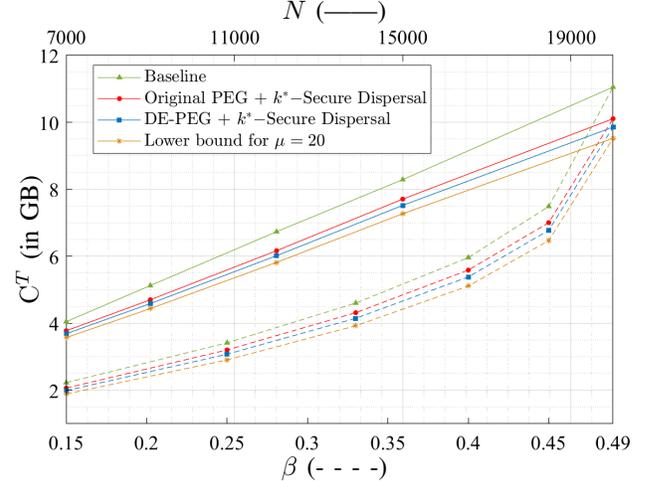

 Table \ref{table:comparison} compares the performance of the $k^*$-secure dispersal protocol and DE-PEG algorithm with the performance achieved by the data availability oracle of \cite{AceD}.
 \debbb{$\mathrm{C}^T_{full}$ is the communication cost associated with} $k_{\min}$ (considering that each node gets the same chunk multiple times and similar to the computation carried out in \cite{AceD}). \debbb{$\mathrm{C}^T_{distinct}$ is the total cost} by considering only distinct chunks (out of the $k_{\min}$ chunks)  at each node (calculated using Monte-Carlo simulations). For our work, we present the baseline and the $k^*$-secure dispersal protocol with $\mu = 20$ (best results in Table \ref{table:mu_variation}) 
 with the PEG and the DE-PEG algorithms. \debbb{From Table \ref{table:comparison}, we see three levels of cost reduction; 1) $\mathrm{C}^T_{distinct}$ to $\mathrm{C}^T_{baseline}$ is due to sampling with replacement, 2)  $\mathrm{C}^T_{baseline}$ to $\mathrm{C}^T_{PEG}$ is  due to using the $k^*$-secure dispersal protocol for $\mu = 20$, and 3) $\mathrm{C}^T_{PEG}$ to $\mathrm{C}^T_{DE-PEG}$ is due to using DE-PEG LDPC codes designed to reduce the total communication cost}. Note that these reductions are for a single 1MB block. Also, for the Oracle of \cite{AceD}, we use $\alpha^* = 0.125$ which is for a rate $\frac{1}{4}$ code but assume the data chunk sizes are the same as a rate $\frac{1}{2}$ code to demonstrate that even in this disadvantageous situation we have a better communication cost.   

In Fig. \ref{fig:combined}, the total communication cost  $\mathrm{C}^T$ is plotted as a function of the number of oracle nodes and adversary fraction. The solid plots show $\mathrm{C}^T$ vs. $N$ at $\beta = 0.49$. We see that at  $N = 15000$, compared to the baseline, there is around $7\%$ reduction in $\mathrm{C}^T$ by using the $k^*$-secure dispersal protocol with $\mu = 20$ and the PEG algorithm, and a $9.3\%$ reduction by using the protocol with $\mu = 20$ and the DE-PEG algorithm. The yellow plot corresponds to the lower bound on $\mathrm{C}^T$ for $\mu = 20$ and is tantamount to a maximum of $13\%$ reduction in $\mathrm{C}^T$ from the baseline.
The dotted plots show $\mathrm{C}^T$ vs. $\beta$ at $N = 20000$. We see a sharp increase in $\mathrm{C}^T$ for higher $\beta$ and as $\beta$ is increased from 0.4 to 0.49, $\mathrm{C}^T$ increases by around $5.1$GB for the baseline, whereas for the PEG and DE-PEG algorithms using the $k^*$-secure dispersal protocol (with $\mu = 20$), $\mathrm{C}^T$ increases by around $4.52$GB and $4.47$GB, respectively. The result indicates that using our methods, the system has to pay less in terms of the total communication cost in order to handle a higher adversary fraction.  
 
In conclusion, we provided a new dispersal protocol and a modification of the PEG algorithm, called the DE-PEG algorithm, that when combined provide a much lower communication cost in the data availability oracle of \cite{AceD} compared to previous schemes. 
\debbbb{Simulation results confirm significant improvement in the communication cost. Additionally, our new constructions are more flexible in terms of system parameters. We are currently investigating other coding techniques, such as Polar codes, in the context of this application.}

\vspace{-0.25cm}
\section*{Acknowledgment}
\vspace{-0.15cm}
The authors acknowledge the Guru Krupa Foundation and NSF-BSF grant no. 2008728 to conduct this research work.

\appendices

\section{Construction of Coded Interleaving Tree }\label{appendix:CIT_construction}

Let the CIT have $l$ layers (except the root), $L_1, L_2, \ldots, L_l$, where $L_l$ is the base layer. For $1 \leq j\leq l$, let $L_j$ have $n_j$ coded symbols, where we use $n_l = M$ in this paper. Let $N_j[i]$, $ 1 \leq i \leq n_j$, be the $i^{th}$ symbol of the $j^{th}$ layer, where $S_j = \{N_j[i]$, $ 1 \leq i \leq \tr n_j\}$ and $P_j = \{N_j[i]$, $ \tr n_j +1 \leq i \leq n_j\}$ are the set of systematic (data) and parity symbols of $L_j$, respectively, where we also write $S_j[i] = N_j[i], 1 \leq i \leq \tr n_j$. $P_j$ is obtained from $S_j$ using a rate $\tr$ systematic LDPC code $H_j$.  In the above CIT, hashes of $q$ coded symbols of every layer are batched (concatenated) together to form a data symbol of its parent layer, where the $n_j$'s satisfy $n_j = \frac{M}{(q\tr)^{l-j}}$, $j = 1,2, \ldots, l$.

Let $s_j = \tr n_j$ and $p_j = (1-\tr)n_j$ denote the number of systematic and parity symbols in $L_j$. Also define $x \bmod {p} := (x)_p$.
The data symbols of  $L_{j-1}$ are formed from the coded symbols of $L_j$ as follows (for $1 < j \leq l$):
\begin{align*}
    S_{j-1}[i] &= N_{j-1}[i] = {\fontfamily{qcr}\selectfont \text{concat}}(\{{\fontfamily{qcr}\selectfont \text{Hash}}(N_{j}[x]) \:\vert \:1\leq x \leq n_j,\\& i = 1+  (x -1)_{s_{j-1}} \}), \quad  1 \leq i \leq s_{j-1},
\end{align*}
\noindent
where ${\fontfamily{qcr}\selectfont \text{Hash}}$ is a hash function (whose output size is $y$) and ${\fontfamily{qcr}\selectfont \text{concat}}$ represents the
string concatenation function. The CIT has a root which consists of $t$ hashes. The CIT allows to create a Proof of Membership (POM) for each base layer coded symbol (which consists of a data and a parity symbol from each intermediate layer of the tree). In particular, the POM of symbol $N_l[i]$ is the set of symbols \{$N_{j}[1 + (i-1)_{s_j}]$, $N_{j}[1 + s_j + (i-1)_{p_j}] \:\vert\: 1 \leq j \leq l-1$\}. The POMs have the \emph{sibling} property \cite{AceD}, i.e., for each layer $j$, $1\leq j < l-1$, the data part of the POM from layer $j$ is the parent of the two symbols of the POM from layer $j+1$. In other words, $N_{j}[1 + (i-1)_{s_j}]$ is the parent of  $N_{j+1}[1 + (i-1)_{s_{j+1}}]$ and $N_{j+1}[1 + s_{j+1} + (i-1)_{p_{j+1}}]$. By parent, we mean that $N_{j}[1 + (i-1)_{s_j}]$ contains the hashes of $N_{j+1}[1 + (i-1)_{s_{j+1}}]$ and $N_{j+1}[1 + s_{j+1} + (i-1)_{p_{j+1}}]$. The POM of a symbol from any intermediate layer $j$ of the tree $1< j < l$ similarly consists of a data and a parity symbol from each layer above layer $j$. In particular, POM of the symbol $N_j[i]$ is the set of symbols \{$N_{j'}[1 + (i-1)_{s_{j'}}]$, $N_{j'}[1 + s_{j'} + (i-1)_{p_{j'}}] \:\vert\: 1 \leq j' < j$\} and they also satisfy the \emph{sibling} property. The POM of a coded symbol is its Merkle proof \cite{Bitcoin} and is used to check the inclusion of the coded symbol in the tree (w.r.t to the CIT root). Note that with the described POM for each symbol, the process of checking the Merkle proof is same as for regular Merkle trees in \cite{Bitcoin}.

Let $X_j$ be the size of one coded chunk of layer $j$ along with its POMs which involves a data and parity symbol from each layer above layer $j$ as defined in Section \ref{sec:design_idea}. Note that the size of each base layer coded chunk is $\frac{b}{\tr M}$, where $b$ is the block size. Thus,
$$X_l = \frac{b}{\tr M} + [(q-1)y + qy ](l-1) = \frac{b}{\tr M} + y(2q-1)(l-1)$$
where the term $(q-1)$ arises due to the fact that of the $q$ hashes present in the data symbol of the POM from layer $j$, $1 \leq j <l-1$, the hash corresponding to the data symbol of the POM from layer $(j+1)$ is not communicated in the POM (since, due to the \emph{sibling} property, it can be calculated by taking a hash of the data symbol of the POM from layer $(j+1)$). Similarly, of the $q$ hashes present in the data symbol of the POM from layer $l-1$, the the hash corresponding to the actual base layer data chunk is not communicated. Thus we only get $(q-1)$ hashes from each layer for the data part in the POMs. Similarly,  $X_j = qy + y(2q-1)(j-1)$, $1\leq j < l$.

\section{Proof of Lemma \ref{lemma:N_upper_bound}}\label{appendix:proof_N_upper_bound}

Let $\rho = \frac{\gamma N k}{M}$, and $x = e^{\rho}$.  The condition $k > \frac{M}{N\gamma}\ln\frac{1}{1-\eta}$ is equivalent to $x > \frac{1}{\Bar{\eta}}$
and the condition $P^{\mathrm{UB}}(\eta,N,M,k,\gamma) \leq p_{th}$ can be simplified to $x^2(v\Bar{\eta} - \Bar{\eta}^2) + (2\Bar{\eta}+v)x -1 \leq 0$. For $N = \frac{M(1-\eta) + \ln(p_{th})}{H_e(\gamma)}$, $v = \Bar{\eta}$
and hence we need $x \leq \frac{1}{3\Bar{\eta}}$ which is not possible for $x > \frac{1}{\Bar{\eta}}$ (note that $\Bar{\eta} >0$). For $N > \frac{M(1-\eta) + \ln(p_{th})}{H_e(\gamma)}$, $v > \Bar{\eta}$ and hence $x^2(v\Bar{\eta} - \Bar{\eta}^2) + (2\Bar{\eta}+v)x -1$ is an upward facing quadratic equation with roots of opposite sign. Since $x$ is always positive,  $x^2(v\Bar{\eta} - \Bar{\eta}^2) + (2\Bar{\eta}+v)x -1 \leq 0$ iff $x \leq x_{\max} = \frac{-(2\Bar{\eta}+v) + \sqrt{8\Bar{\eta}v+v^2}}{2\Bar{\eta}(v-\Bar{\eta})}$, where $x_{\max}$ is the positive root of the quadratic equation. However, a quick algebraic check would reveal that $x_{\max} < \frac{1}{\Bar{\eta}}$ for $v > \Bar{\eta}$ and hence there is no feasible $x$ which satisfies $x^2(v\Bar{\eta} - \Bar{\eta}^2) + (2\Bar{\eta}+v)x -1 \leq 0$. Thus, for $N \geq \frac{M(1-\eta) + \ln(p_{th})}{H_e(\gamma)}$, $P^{\mathrm{UB}}(\eta,N,M,k,\gamma) > p_{th}$  $\forall k > \frac{M}{N\gamma}\ln\frac{1}{1-\eta}$.

For $N < \frac{M(1-\eta) + \ln(p_{th})}{H_e(\gamma)}$, $v < \Bar{\eta}$. In this case $x^2(v\Bar{\eta} - \Bar{\eta}^2) + (2\Bar{\eta}+v)x -1$ is a downward facing quadratic equation with roots
$x_{\max} = \frac{-(2\Bar{\eta}+v) - \sqrt{8\Bar{\eta}v+v^2}}{2\Bar{\eta}(v-\Bar{\eta})}$ and $x_{\min} = \frac{-(2\Bar{\eta}+v) + \sqrt{8\Bar{\eta}v+v^2}}{2\Bar{\eta}(v-\Bar{\eta})}$ satisfying $x_{\min} < \frac{1}{\Bar{\eta}} < x_{\max}$. In this situation, $x^2(v\Bar{\eta} - \Bar{\eta}^2) + (2\Bar{\eta}+v)x -1 \leq 0$ iff $x \geq x_{\max}$ which is equivalent to $k \geq \frac{M}{N\gamma}\ln\left(\frac{-(2\Bar{\eta}+v) - \sqrt{8\Bar{\eta}v + v^2}}{2\Bar{\eta}(v-\Bar{\eta})}\right)$.

\section{Proof of Lemma \ref{lemma:coupon_group}}\label{appendix:coupon_group}
We use the following result from \cite{CouponGroup}. 
\begin{lemma} (\cite{CouponGroup})
Let $S$ be a set of $s$ elements and let $A \subseteq S, \vert A\vert = l$. From $S$, let $T$ subsets of size $m$ be drawn with replacement, each subset drawn uniformly at random from all subsets of size $m$ of $S$. Let $X_{T}(A)$ be the number of distinct elements of the set $A$ contained in the above $T$ drawings. Then
\begin{align*}
    \mathrm{Prob}(X_{T}(A) \leq n) &:= \chi(n,l,s,T,m)\\= \sum_{j=0}^{n}(-1)^{n-j}{l \choose j}&{l - j - 1 \choose l-n-1}\left[\frac{{s-l+j \choose m}}{{s \choose m}}\right]^T.
\end{align*}

\end{lemma}

\noindent
Now, following in a manner similar to \cite[Appendix A]{AceD}
\begin{align*}
    \mathrm{Prob}(\tC \text{ is not } \mu\text{-SS-valid})
    \text{\hspace{4.2cm}}
\end{align*}    
\vspace{-0.65cm}
\begin{align*}
         \text{\hspace{0.8cm}}&= \mathrm{Prob}(\exists S \text{ such that } \vert S \vert = \gamma N, \vert \cup_{i \in S}A_i \vert \leq M - \mu) \\
    &\leq  \sum_{S \subseteq [M] : \vert S \vert = \gamma N}\mathrm{Prob}(\vert \cup_{i \in S}A_i \vert \leq M - \mu)\\
    &= \sum_{S \subseteq [M] : \vert S \vert = \gamma N}\chi(M - \mu, M,M,\gamma N,k)\\
        &= {N \choose \gamma N}\chi(M - \mu, M,M,\gamma N,k)\\
    &\leq e^{NH_e(\gamma)}\chi(M - \mu, M,M,\gamma N,k)\\
    &= e^{NH_e(\gamma)}P_{f}%
\end{align*}    
where similar to \cite{AceD}, we have used the fact that ${N \choose \gamma N} \leq e^{N H_e(\gamma)}$ and $\mathrm{Prob}(\vert \cup_{i \in S}A_i \vert \leq M - \mu) = \mathrm{Prob}(X_{T}(A) \leq n)$ when $S = A$, $l = s = M$, $n = M - \mu$, $m=k$ and $T = \gamma N$. 

\section{Proof of Lemma \ref{lemma:dispersal_availability}}\label{appendix:dispersal_availability}

 In the secure phase of Dispersal protocol \ref{dispersal_protocol}, for each layer $j$, $1 \leq j \leq l$, all stopping sets (of $H_j$) of sizes $< (n_j - \left \lceil{\left(\frac{M - \mu +1}{M}\right)n_j}\right \rceil + 1)$ are securely dispersed. Hence a peeling decoder will never fail to decode layer $j$, $1 \leq j \leq l$ due to these stopping sets. 

Furthermore, the valid phase of Dispersal protocol \ref{dispersal_protocol} ensures that every $\gamma$ fraction of the oracle nodes have at least $M - \mu + 1$ distinct base layer coded chunks with probability $\geq 1 - p_{th}$. Thus, due to the repetition property described in Section \ref{section:preliminaries}, this ensures that for a given layer $j$, $1 \leq j < l$, every $\gamma$ fraction of the oracle nodes have has at least $\frac{M - \mu +1}{M}$ fraction of distinct coded chunk, or at least  $\left \lceil{\left(\frac{M - \mu +1}{M}\right)n_j}\right \rceil$ distinct coded chunks. Thus,  with probability $\geq 1 - p_{th}$, the dispersal protocol is $(n_j - \left \lceil{\left(\frac{M - \mu +1}{M}\right)n_j}\right \rceil + 1)$-SS-valid for each layer $j$, $1\leq j\leq l$.

Since the CIT root is committed only when $\gamma + \beta$ fraction of the oracle nodes vote that they received correct coded chunks, this implies that at least $\gamma$ fraction of honest oracle nodes have received correct coded chunks. Now, since with probability $\geq 1 - p_{th}$, the dispersal protocol is $(n_j - \left \lceil{\left(\frac{M - \mu +1}{M}\right)n_j}\right \rceil + 1)$-SS-valid for each layer $j$,  a peeling decoder can successfully decoder layer $j$ for all stopping sets of size $\geq (n_j - \left \lceil{\left(\frac{M - \mu +1}{M}\right)n_j}\right \rceil + 1)$ by downloading the coded chunks from the above honest $\gamma$ fraction of oracle nodes who voted that they received correct coded chunks.

Combining the above two situations, the decoder can decode the entire CIT if the block is committed. Hence Dispersal Protocol \ref{dispersal_protocol} guarantees availability with probability $\geq 1 - p_{th}$.

\end{document}